\documentclass[aps,twocolumn,preprintnumbers,showpacs]{revtex4}

\usepackage{epsfig}

\usepackage{bbm}

\newcommand{\lbfig}[1]{\refstepcounter{fig} \label{#1} }
\newcounter{fig}

\newcommand{\nc}{\newcommand}
\nc{\be}{\begin{equation}}
\nc{\ee}{\end{equation}}
\nc{\bea}{\begin{eqnarray}}
\nc{\eea}{\end{eqnarray}}
\nc{\bi}[1]{\bibitem{#1}}
\nc{\lsim}{\mbox{\raisebox{-.6ex}{~$\stackrel{<}{\sim}$~}}}
\nc{\gsim}{\mbox{\raisebox{-.6ex}{~$\stackrel{>}{\sim}$~}}}

\begin{document}

\preprint{HD-THEP-02-39}

\vskip 0.2in

\title{Particle number in kinetic theory}

\author{Bj\"orn Garbrecht$^*$, Tomislav Prokopec$^*$
                       and 
                   Michael G. Schmidt
       }
\email[]{B.Garbrecht@thphys.uni-heidelberg.de}
\email[]{T.Prokopec@thphys.uni-heidelberg.de}
\email[]{M.G.Schmidt@thphys.uni-heidelberg.de}

\affiliation{Institut f\"ur Theoretische Physik, Heidelberg University,
             Philosophenweg 16, D-69120 Heidelberg, Germany}


\begin{abstract}
We provide a derivation for the particle number densities on phase space for
scalar and fermionic fields in terms of Wigner functions.
Our expressions satisfy the desired properties:
for bosons the particle number is positive, 
for fermions it lies in the interval between {\it zero} and {\it one},
and both are consistent with thermal field theory.
As applications we consider the Bunch-Davies vacuum and 
fermionic preheating after inflation.
\end{abstract}

\pacs{98.80.Cq, 05.60.Gg, 04.62.+v, 98.80.-k }

\maketitle

%
%

\section{Introduction}
The notion of particles is very intuitive, and at the classical 
level, in statistical physics, the dynamics is very successfully
described by the classical Boltzmann equation for particle densities
in phase space. In quantum physics however, the uncertainty principle
seems to prohibit the use of phase space densities, and they are replaced
by their closest analogues, the Wigner 
functions~\cite{Wigner:1932,HilleryOConnellScullyWigner:1983}.
Yet, strictly speaking they 
can neither be interpreted as particle
numbers nor as probability distributions on phase space, since they may
aquire negative values. Attempts have been made to define particle
number in relativistic quantum kinetic 
theory~\cite{KlugerMottolaEisenberg:1998},
but so far there exists no result that would be applicable
to general situations.

In spite of those difficulties, the dynamics of quantum fields and
particle numbers in the presence of temporally varying background fields
has been extensively studied and is well 
understood~\cite{Parker:1969,MamaevMostepanenkoFrolov:1976,ChungKolbRiottoTkachev:1999}.
The particle number
operator can be calculated by a Bogolyubov transformation rotating the
Fock space to a new basis, which mixes positive and negative
frequency solutions.

In the analysis presented in this paper we show that the Wigner function,
which we take here as an expectation value with respect to the ground state
of the original basis, provides the necessary 
information about the rotated basis to calculate the particle number
produced by the coupling to time-dependent external fields.

\section{Scalars}
\subsection{Scalar kinetic equations}
 As the first model case, we consider a massive scalar field minimally
coupled to gravity, such that in a conformal space-time,
with the metric of the form $g_{\mu\nu}=a^2\eta_{\mu\nu}$,
the Lagrangean is given by 
\begin{equation}
\sqrt{-g}\mathcal{L}_\Phi = \frac{1}{2}a^2\eta^{\mu\nu}
                       (\partial_{\mu}\Phi)(\partial_{\nu}\Phi)
                     - \frac{1}{2}a^4m_\phi^{2}\Phi^{2}
,
\label{Lagrangian:scalar}
\end{equation}
where $\eta^{\mu\nu} = {\rm diag}[1,-1,-1,-1]$ is the Minkowski (flat) 
metric, and $a=a(\eta)$ is the scale factor. For example, in inflation
$a = -1/(H\eta)$ ($\eta<0$), while in radiation-matter era, 
$a = a_r\eta+a_m\eta^2$. Here $\eta$ denotes conformal time,
$a_r$ and $a_m$ are constants. 

We quantize the theory~(\ref{Lagrangian:scalar}) by promoting
$\Phi(x)$ to an operator,
\begin{eqnarray}
\Phi(x) \equiv \frac{\mathbf\varphi}a
= \frac{1}{aV} \sum_{\mathbf{k}}
             e^{-i\mathbf{k}\cdot\mathbf{x}}
          \Bigl(\varphi_{\mathbf{k}}(\eta)a_{\mathbf{k}}
        + \varphi^{*}_{-\mathbf{k}}(\eta)a^{\dagger}_{-\mathbf{k}}\Bigr)
,
\nonumber
\end{eqnarray}
where $V$ denotes the comoving volume. 
The mode functions obey the Klein-Gordon equation
\begin{equation}
\bigl(\partial^{2}_{\eta}+\omega_{\mathbf{k}}^2-{a''}/{a}
\bigr)\varphi_{\mathbf{k}}
    = 0,
\label{Klein-Gordon}
\end{equation}
where $^\prime \equiv d/d\eta$,
$\omega_{\mathbf{k}}^2 = \mathbf{k}^{2} + a^{2}m_\phi^{2}(\eta)$
defines the single particle (comoving) energy, and we take for the Wronskian
\begin{equation}
  \varphi_{\mathbf{k}}^*\varphi_{\mathbf{k}}'
 -{\varphi_{\mathbf{k}}^*}'\varphi_{\mathbf{k}} = i
.
\label{Wronskian}
\end{equation}
%
Throughout this paper we assume that the modes 
$\varphi_{\mathbf{k}} =\varphi_k$ ($k \equiv |\mathbf{k}|$) are 
homogeneous, which is justified when the mass is varying slowly in space,
such that we can ignore its gradients.
The field $\varphi = a\Phi$ obeys the canonical commutation relation,
\begin{equation}
[\varphi(\mathbf{x},\eta),\partial_\eta\varphi(\mathbf{x'},\eta)] 
= i\delta^3(\mathbf{x}-\mathbf{x'}),
\label{canonical-commutation}
\end{equation}
\noindent
which implies $[a_\mathbf{k},a^\dagger_\mathbf{k'}] 
 = \delta_{\mathbf{k},\mathbf{k'}}$.

The fundamental quantity of quantum kinetic theory is the two-point
Wightman function, which we here write for the ground state $|0\rangle$  
annihilated by $a_{\mathbf{k}}$, $a_{\mathbf{k}}|0\rangle = 0$.
With the rescaling suitable for conformal space-times, it reads
\begin{equation}
i\bar{G}^{<}(u,v)\equiv a(u)iG^{<}(u,v)a(v)
                    =  \left\langle 0|\varphi(v)\varphi(u)|0\right\rangle
,
\label{Wightman-function:bosons}
\end{equation}
and its Wigner transform is defined as 
\begin{eqnarray}
iG^{<}(k,x)=\int d^{4}r {\rm e}^{ik\cdot r}iG^{<}(x+r/2,x-r/2)
\,,
\nonumber
\end{eqnarray}
which satisfies the  Klein-Gordon 
equation~\cite{ZhuangHeinz:1998,BodekerKainulainenProkopec:unpub}
%
\begin{eqnarray}
\Big(-ik_{0}\partial_{\eta}+\frac{1}{4}\partial^{2}_{\eta}-k^{2}
     +\bar{m}_\phi^{2}(\eta)
      {\rm e}^{-\frac{i}{2}\overleftarrow{\partial}_{\eta}\partial_{k_{0}}}
\Big)i\bar{G}^{<} 
  = 0
,\quad
\label{ke+ce}
\end{eqnarray}
where 
$
 \bar{m}_\phi^{2} = a^{2}m_\phi^{2} - a''/a.
$
It is then useful to define the $n$-th moments of the Wigner function,
\begin{equation}
    f_{n}(\mathbf{k},x) 
  \equiv \int\frac{dk_{0}}{2\pi} k^{n}_{0}i\bar{G}^{<}(k,x)
.
\label{moments:bosons}
\end{equation}
Taking the 1st (0th) moment of the imaginary 
(real) part of Eq.~(\ref{ke+ce}) 
yields~\cite{ZhuangHeinz:1998,BodekerKainulainenProkopec:unpub}
\begin{eqnarray}
f'_2 - \frac{1}{2}  {(\bar m_\phi^2)}' f_{0} = 0
,\qquad
\frac{1}{4}f_{0}'' - f_{2} + \bar{\omega}_{\mathbf{k}}^{2} f_{0} = 0
,
\label{moments}
\end{eqnarray}
with $\bar\omega_{\mathbf{k}}^2 = \mathbf{k}^2+\bar{m}_\phi^{2}$.
Eliminating $f_{2}$ from ~(\ref{moments}) 
yields~\cite{BodekerKainulainenProkopec:unpub}
\begin{equation}
    f_{0}'''
  + 4\bar \omega_{\mathbf{k}}^2f_{0}'
  + 2(\bar \omega_{\mathbf{k}}^2)' f_{0}
  = 0
.
\label{bo2P}
\end{equation}
This can be integrated once to give
\begin{equation}
   \bar\omega_{\mathbf{k}}^{2}f_0^2 + \frac{1}{2}f_0{''}f_0
 - \frac{1}{4}{f_0'}^2
 = \frac{1}{4}
\,,
\label{eom:f02}
\end{equation}
where the integration constant is obtained by making use of
$f_0 = |\varphi_{\mathbf{k}}|^2$ ({\it cf.} Eq.~(\ref{dotg}) below),
Eq.~(\ref{Klein-Gordon}) and the Wronskian~(\ref{Wronskian}).

\subsection{Bogolyubov transformation}
\label{sec:Bogolyubov transformation}
The Hamiltonian density corresponding to 
the Lagrangean~(\ref{Lagrangian:scalar}) reads
\begin{eqnarray}
H &=& \frac{1}{2V}\sum_{\mathbf{k}}\left\{
      \Omega_{\mathbf{k}}(a_{\mathbf{k}}a^{\dagger}_{\mathbf{k}}
   +  a^{\dagger}_{\mathbf{k}}a_{\mathbf{k}})
   +  (\Lambda_{\mathbf{k}} a_{\mathbf{k}}a_{-\mathbf{k}} 
   + {\rm h.c.})
                                   \right\}
\nonumber\\
\Omega_{\mathbf{k}}
    &=& \left|\varphi_{{k}}' - (a'/a)\varphi_{{k}}
        \right|^2
     + \omega_{\mathbf{k}}^2 \left|\varphi_{{k}}\right|^2
\nonumber\\
\Lambda_{\mathbf{k}}
   &=& \Bigl(\varphi_{{k}}'-\frac{a'}{a}\varphi_{{k}}\Bigr)^2
    +  \omega_{\mathbf{k}}^2\varphi_{{k}}^2
.\qquad
\label{Hamiltonian:bosons}
\end{eqnarray}

Consider now the homogeneous Bogolyubov transformation
\begin{equation}
  \left(\begin{array}{c}
            \hat{a}_{\mathbf{k}}\\
            \hat{a}^{\dagger}_{-\mathbf{k}}
        \end{array}\right)
 = \left(\begin{array}{cc}
             \alpha_{{k}}&\beta^{*}_{{k}}\\
             \beta_{{k}}&\alpha^{*}_{{k}}
         \end{array}
   \right)
   \left(\begin{array}{c}
             a_{\mathbf{k}}\\
             a^{\dagger}_{-\mathbf{k}}
         \end{array}
   \right)
,
\label{Bogolyubov-transformation}
\end{equation}
with the norm
\begin{equation}
     |\alpha_k|^2 - |\beta_k|^2 = 1
\,,
\label{alpha-beta:norm}
\end{equation}
upon which $\Lambda_{\mathbf{k}}$ and $\Omega_{\mathbf{k}}$ transform as 
\begin{eqnarray}
 \Lambda^\prime_{\mathbf{k}} &=& - 2\alpha_k^* \beta_k \Omega_{\mathbf{k}}
                      + (\alpha_k^*)^2 \Lambda_{\mathbf{k}}
                      + \beta_k^2 \Lambda_{\mathbf{k}}^* 
\label{lambda}
\\
 \Omega^\prime_{\mathbf{k}} &=&  (|\alpha_k|^2+|\beta_k|^2) \Omega_{\mathbf{k}}
                      - \alpha_k^*\beta_k^* \Lambda_{\mathbf{k}}
                      - \alpha_k\beta_k \Lambda_{\mathbf{k}}^* 
\label{omega}
\,.
\end{eqnarray}
In terms of real and imaginary parts, these equations
can be recast as
\begin{eqnarray}
2 |\alpha_k|  |\beta_k| \Omega_{\mathbf{k}}
+ |\Lambda^\prime_{\mathbf{k}}| \cos(\phi_\lambda\!+\!\phi_\alpha\!-\!\phi_\beta)
\nonumber\\
    - (|\alpha_k|^2\!\! +\! |\beta_k|^2) |\Lambda_{\mathbf{k}}|
                             \cos(\phi_\Lambda\!-\!\phi_\alpha\!-\!\phi_\beta)
\!\!\!&=&\!\!\! 0,
\label{lambda:1}
\\
\!\!\!
 |\Lambda^\prime_{\mathbf{k}}| \sin(\phi_{\Lambda^\prime}\!+\!\phi_\alpha\!-\!\phi_\beta)
   \! -\!  |\Lambda_{\mathbf{k}}|\sin(\phi_\Lambda\!-\!\phi_\alpha\!-\!\phi_\beta\!\!)\! &=&\!\!\! 0,
\label{lambda:2}
\\
 \Omega^\prime_{\mathbf{k}} 
 \!-\!  (|\alpha_k|^2\!\!+\!|\beta_k|^2) \Omega_{\mathbf{k}}\!
\nonumber\\
 + 2|\alpha_k|  |\beta_k|  |\Lambda_{\mathbf{k}}|
                      \cos(\phi_\Lambda\!-\!\phi_\alpha\!-\!\phi_\beta) 
 \!\!\!  &=&\!\!\! 0
\label{omega:1}
,
\end{eqnarray}
with $|\alpha_k| = \sqrt{1+|\beta_k|^2}$, and where we have introduced the
phases
\begin{eqnarray}
\Lambda^\prime_{\mathbf{k}}&=|\Lambda^\prime_{\mathbf{k}}|
\exp\left(i\phi_{\Lambda^\prime}\right)\, , \quad
\Lambda_{\mathbf{k}}&=|\Lambda_{\mathbf{k}}|
\exp\left(i\phi_{\Lambda}\right)\\
\alpha_k&=|\alpha_k|
\exp\left(i\phi_\alpha\right)\, , \quad
\beta_k&=|\beta_k|
\exp\left(i\phi_\beta\right).
\end{eqnarray}
Eqs.~(\ref{lambda:1}) 
and~(\ref{omega:1}) can be combined to give
\begin{equation}
  \cos(\phi_\lambda+\phi_\alpha-\phi_\beta) 
        =  \frac{(|\alpha_k|^2+|\beta_k|^2) \Omega_{\mathbf{k}}
        - \Omega^\prime_{\mathbf{k}}}
                {2|\alpha_k|\,|\beta_k|\,|\Lambda_{\mathbf{k}}|}
\label{angles:1}
\,,
\end{equation}
while (\ref{omega:1}) yields an expression for 
$\cos(\phi_\Lambda-\phi_\alpha-\phi_\beta)$. Upon squaring Eq.~(\ref{lambda:2})
and making use of $\sin^2(\zeta) = 1- \cos^2(\zeta)$, we find that 
\begin{equation}
 \Omega_{\mathbf{k}}^2 - |\Lambda_{\mathbf{k}}|^2 
                =  {\Omega^\prime_{\mathbf{k}}}^2 
                -  {|\Lambda^\prime_{\mathbf{k}}|}^2
\label{constraint:BT}
\end{equation}
is an invariant of 
the Bogolyubov transformations~(\ref{Bogolyubov-transformation}).

 Next, we solve~(\ref{omega:1}) for $n_{\mathbf{k}} \equiv |\beta_k|^2$ 
to find
\begin{equation}
  n_{\mathbf{k}\pm} \!=\! \frac{\Omega_{\mathbf{k}}\Omega^\prime_{\mathbf{k}} 
                           \pm\sqrt{|\Lambda_{\mathbf{k}}|^2 x^2
                               ({\Omega^\prime_{\mathbf{k}}}^2
                                    - \Omega_{\mathbf{k}}^2
                                          + |\Lambda_{\mathbf{k}}|^2 x^2)}} 
                       {2(\Omega_{\mathbf{k}}^2 - |\Lambda_{\mathbf{k}}|^2 x^2)}
                  \!- \! \frac 12
\label{nk:x}
\end{equation}
where $x\equiv\cos(\varphi_\Lambda-\varphi_\alpha-\varphi_\beta)$.
Upon extremizing this with respect to $x^2$, one can show that 
a maximum is formally reached for 
$x^2_{\rm max} = \Omega_{\mathbf{k}}^2/|\Lambda_{\mathbf{k}}|^2$,
which must be greater than one if
the Hamiltonian~(\ref{Hamiltonian:bosons}) is to be diagonalizable. 
Taking account of $x^2 \leq 1$, one finds that the maximum for
$n_{\mathbf{k}\pm}$ is reached when $x^2 = 1$, for which 
\begin{equation}
 n_{\mathbf{k}\pm} = \frac{\Omega_{\mathbf{k}}\sqrt{\Omega_{\mathbf{k}}^2 
                                     -   |\Lambda_{\mathbf{k}}|^2
                                     +   |{\Lambda^\prime_{\mathbf{k}}}|^2}
                    \pm|\Lambda_{\mathbf{k}}|\, |\Lambda^\prime_{\mathbf{k}}|}
                   {2(\Omega_{\mathbf{k}}^2 - |\Lambda_{\mathbf{k}}|^2)}
               - \frac 12
\,.
\end{equation}
Since  $n_{\mathbf{k} -} = 0$ when
 $|\Lambda^\prime_{\mathbf{k}}| = |\Lambda_{\mathbf{k}}|$,
the physical branch corresponds to $n_{\mathbf{k}} = n_{\mathbf{k} -}$.
Furthermore, when considered as a function 
of $|\Lambda^\prime_{\mathbf{k}}|$, 
$n_{\mathbf{k}} \equiv n_{\mathbf{k} - }$ monotonously increases 
as $|\Lambda^\prime_{\mathbf{k}}|$ decreases, reaching a maximum when 
$|\Lambda^\prime_{\mathbf{k}}| = 0$ (see figure~\ref{figure 0}), 
for which the particle number is
\begin{equation}
 n_{\mathbf{k}} 
     = \langle 0|\hat{a}^{\dagger}_{\mathbf{k}}\hat{a}_{\mathbf{k}}|0\rangle
     = \frac{\Omega_{\mathbf{k}}}{2\omega_{\mathbf{k}}}
     - \frac 12
\,,
\label{particle-number:bosons}
\end{equation}
where $\omega_{\mathbf{k}} = \sqrt{\mathbf{k}^2 + a^2 m_\phi^2}$. 
\begin{figure}[htbp]
\begin{center}
\epsfig{file=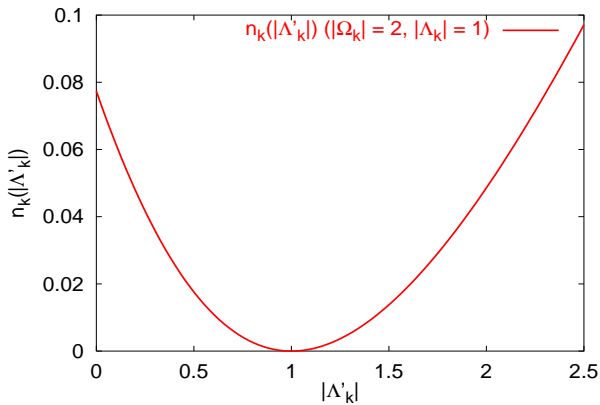, height=2.2in,width=3.3in}
\end{center}
\vskip -0.15in
\lbfig{figure 0}
\caption[fig0]{%
\small
  Particle number $n_{\mathbf{k}}$ as a function of 
$|\Lambda^\prime_{\mathbf{k}}|$ 
for $|\Omega_{\mathbf{k}}| = 2$,  $|\Lambda_{\mathbf{k}}| = 1$.
Provided $|\Lambda^\prime_{\mathbf{k}}| \leq |\Lambda_{\mathbf{k}}|$, 
$n_{\mathbf{k}}$ maximizes at $|\Lambda^\prime_{\mathbf{k}}| = 0$.
}
\end{figure}
This definition, which corresponds to the (constrained) maximum
possible particle number a detector can observe, 
we shall use as our definition for particle number on phase space.
Moreover, note that, in terms of thus transformed creation and annihilation
operators $\hat{a}^{\dagger}_{\mathbf{k}}$ and $\hat{a}_{\mathbf{k}}$,
the Hamiltonian is diagonal
\begin{eqnarray}
H  = \frac{1}{2V} \sum_{\mathbf{k}}
      \omega_{\mathbf{k}}(\hat a_{\mathbf{k}}\hat a^{\dagger}_{\mathbf{k}}
   +  \hat a^{\dagger}_{\mathbf{k}}\hat a_{\mathbf{k}})
,\quad
      [\hat{a}_{\mathbf{k}},\hat{a}^{\dagger}_{\mathbf{k'}}]
   = \delta_{\mathbf{k},\mathbf{k'}},
\label{hamiltonian:diag}
\end{eqnarray}
such that our definition agrees with the one advocated, for example,
in Refs.~\cite{Parker:1969,KofmanLindeStarobinsky:1997}.

\vskip 0.1in

\subsection{Particle number in scalar kinetic theory}

It is now a simple matter to calculate the particle number
in terms of Wigner functions. Making use of~(\ref{Wightman-function:bosons})
and~(\ref{moments:bosons}) we find
\begin{eqnarray}
|\varphi_{{k}}|^2 &=& f_0
,\qquad
|{\varphi'}_{{k}}|^{2} = \frac{1}{2}f_0'' +\bar\omega_{\mathbf{k}}^{2}f_0
\label{ddotg}
,
\label{dotg}
\end{eqnarray}
from which it follows
\begin{equation}
\Omega_{\mathbf{k}}
   =  2\left(\omega_{\mathbf{k}}^2 f_0 + \frac{1}{4}f''_0\right)
   - \frac{d}{d\eta}\bigg(\frac{a'}{a}f_0\bigg)
.
\label{Omega:kinetic-theory}
\end{equation}
We then insert~(\ref{Omega:kinetic-theory}) 
into~(\ref{particle-number:bosons}) to get   
\begin{eqnarray}
      n_{\mathbf{k}}
  &=& \omega_{\mathbf{k}} f_0 + \frac{1}{4\omega_{\mathbf{k}}} f_0''  -  \frac{1}{2}
   -  \frac{1}{2\omega_{\mathbf{k}}}\frac{d}{d\eta}\bigg(\frac{a'}{a}f_0\bigg)
.
\label{particle-number:scalars:kin}
\end{eqnarray}
This is our main result for scalars, which is 
positive, simply because
$n_{\mathbf{k}} \equiv |\beta_k|^2\geq 0$ (see Eq.~(\ref{particle-number:bosons})).
Eq.~(\ref{particle-number:scalars:kin}) 
is of course not a unique definition of particle number. Indeed, any 
Bogolyubov 
transformation~(\ref{Bogolyubov-transformation}-\ref{alpha-beta:norm})
corresponds to some particle number definition. 
Our definition~(\ref{particle-number:scalars:kin}) is however 
the special one, in that is correponds to the detector with the best 
possible resolution, {\it i.e.} which measures the maximum number of
particles, as we showed in section~\ref{sec:Bogolyubov transformation}.

 We now apply~(\ref{particle-number:scalars:kin}) to the 
Chernikov-Tagirov~\cite{ChernikovTagirov:1968} 
(Bunch-Davies~\cite{BunchDavies:1978}) vacuum, 
\begin{equation}
 \varphi_k = \frac{1}{\sqrt{2k}}\Big(1-\frac{i}{k\eta}\Big){\rm e}^{-ik\eta}
,
\label{BunchDavies}
\end{equation}
which corresponds to the mode functions of a minimally coupled massless
scalar field in de Sitter inflation, $a=-1/H\eta$ 
({\it cf.} Eq.~(\ref{Klein-Gordon})), for which 
$
 f_0 = ({2k})^{-1}\big(1+{1}/{(k\eta)^2}\big)
,
$
leading to the particle number
%
\begin{eqnarray}
      n_{\mathbf{k}} &=& \frac{1}{4k^2\eta^2}
                      =  a^2\Big(\frac{H}{2k}\Big)^2
.
\label{particle-number:BunchDavies}
\end{eqnarray}
This is to be compared with~\cite{Mijic:1998},
which finds $n_{\mathbf{k}} \sim (-k\eta)^{-3}$ ($-k\eta\ll 1$).
We suspect that the difference is due to the approximate 
method used in~\cite{Mijic:1998}.
On the other hand, when considering the transition from de Sitter inflation
to radiation, one finds that 
the spectrum $n_{\mathbf{k}} \sim (-k\eta_0)^{-4}$ ($-k\eta_0)\gg 1$
($\eta_0$ denotes conformal time at the end of inflation)
is produced~\cite{Ford:1986}.

As a consistency check, we now apply~(\ref{particle-number:scalars:kin})
to thermal equilibrium, where the Wigner function is
({\it cf.} Ref.~\cite{LeBellac:1996})
\begin{eqnarray}
   iG^<
  = 2\pi\textnormal{sign}(k_{0})\delta(k^{2}-m_\phi^{2})
    \frac{1}{{\rm e}^{\beta k_{0}}-1}
.
\label{G< thermal}
\end{eqnarray}
By making use of~(\ref{moments:bosons}) 
and~(\ref{particle-number:scalars:kin}) we obtain
the standard Bose-Einstein distribution,
$
 n_{\mathbf{k}} = 1/({\rm e}^{\beta \omega_{\mathbf{k}}}-1).
$

 Recently, an interesting particle number definition has been
proposed in Ref.~\cite{AartsBerges:2001}, according to which 
(expanding space-times are not considered):

\begin{equation}
 \Big(\tilde n_{\mathbf{k}}+ \frac 12\Big)^2 = |\phi_k|^2 \, |\phi^\prime_k|^2
             = f_0\Big(\frac 12 f_0^{\prime\prime} 
             + \bar\omega^2_{\mathbf{k}}f_0\Big)
\,.
\label{n:AartsBerges}
\end{equation}
Note that in adiabatic domain, in which 
 $f_0^{\prime\prime} \rightarrow 0$,
Eqs.~(\ref{n:AartsBerges}) and~(\ref{particle-number:scalars:kin}) 
both reduce to $n_{\mathbf{k}} \rightarrow \omega_{\mathbf{k}} f_0 - 1/2$,
such that for example in thermal equilibrium of a free scalar
theory~(\ref{G< thermal}), both definitions yield the Bose-Einstein distribution.
According to the authors of~\cite{AartsBerges:2001},
the definition~(\ref{n:AartsBerges}) should be applicable to general situations
(whenever there is a reasonably accurate quasiparticle picture of the plasma),
and it is obtained as a consistency requirement on 
the energy conservation and quasiparticle current relation, respectively,
\begin{equation}
   \frac{\omega_{\mathbf{k}}^2}{2} |\phi_k|^2 
     + \frac{1}{2} |\phi^\prime_k|^2 
     = \omega_{\mathbf{k}}\Big(\frac 12 + \tilde n_{\mathbf{k}}\Big)
\,,\quad
     \omega_{\mathbf{k}} |\phi_k|^2 = \frac 12 + \tilde n_{\mathbf{k}}
\,.
\label{consistency relations}
\end{equation}
The consistency is reached when the kinetic and potential energies are equal,
in which case a generalized quasiparticle energy is given by,
$\omega_{\mathbf{k}}^2 = |\phi^\prime_k|^2 /|\phi_k|^2$.

In order to make a nontrivial comparison, consider now 
a pure state of a scalar theory interacting only weakly with a classical
background field (which can be described by a time dependent mass term). 
The WKB form for the mode functions can be recast as
\begin{eqnarray}
   \phi_{{k}}\!\!\! & = &\!\!\!
   \frac{1}{\sqrt{2\epsilon_{\mathbf{k}}}} 
          \big({\alpha_0
               {\rm e}^{-i\!\!\int^\eta \epsilon_{\mathbf{k}}(\eta^\prime)d\eta^\prime}
            \!\! + \beta_0
               {\rm e}^{i\!\!\int^\eta\epsilon_{\mathbf{k}}(\eta^\prime)d\eta^\prime}}
           \big)
\label{WKB state}
\\
  {\phi}_{{k}}'\!\!\! & = & \!\!\!
 - i\sqrt{\frac{\epsilon_{\mathbf{k}}}{2}}
   (\alpha_0{\rm e}^{-i\!\!\int^\eta\epsilon_{\mathbf{k}}(\eta^\prime)d\eta^\prime}
  \!\!-\beta_0{\rm e}^{i\!\!\int^\eta\epsilon_{\mathbf{k}}(\eta^\prime)d\eta^\prime})
 \! - \frac 12 \frac{\epsilon_{\mathbf{k}}^\prime}{\epsilon_{\mathbf{k}}} \phi_k
\nonumber
\,,
\end{eqnarray}
where $\epsilon_{\mathbf{k}}$ satisfies,
$\epsilon_{\mathbf{k}}^2 = \omega_{\mathbf{k}}^2  
   - (1/2)\epsilon_{\mathbf{k}}^{\prime\prime}/\epsilon_{\mathbf{k}} 
   +(3/4)(\epsilon_{\mathbf{k}}^{\prime\prime}/\epsilon_{\mathbf{k}})^2$. 
In a free theory $|\alpha_0|^2 - |\beta_0|^2 $ is conserved, and it is 
usually normalised to {\it one}. In an interacting theory however, 
the single particle description breaks down, and consequently
 $|\alpha_0|^2-|\beta_0|^2 $ is not conserved. For the purpose
of this example, we assume that the interactions are weak enough, such that
$|\alpha_0|^2-|\beta_0|^2 $ is changing sufficiently slow, and the subsequent
discussion applies.
In the adiabatic limit $\epsilon_{\mathbf{k}}\rightarrow
               \omega_{\mathbf{k}}\rightarrow {\rm constant}$,
the particle number~(\ref{particle-number:bosons}) 
and~(\ref{particle-number:scalars:kin}) of the state~(\ref{WKB state}) is
simply $n^{(0)}_{\mathbf{k}} = |\beta_0|^2$.

 On the other hand, when applied to the state~(\ref{WKB state}), 
the definition~(\ref{n:AartsBerges}) yields an oscillating particle number
even in adiabatic regime,
\begin{equation}
 \Big(\tilde n_{\mathbf{k}}+ \frac 12\Big)^2 \approx
   \frac 14 + (1+|\beta_0|^2)\,|\beta_0|^2
          \sin^2(2\epsilon_{\mathbf{k}}\eta -\chi_\alpha+\chi_\beta)
\,,
\label{n:AartsBerges:2}
\end{equation}
where  $\alpha_0 = |\alpha_0| e^{i\chi_\alpha},
        \beta_0 = |\beta_0| e^{i\chi_\beta}$, which is positive and 
bounded from above by $\tilde n_{\mathbf{k}}\leq 
       |\beta_0|^2 \equiv n_{\mathbf{k}}^{(0)}$
\footnote{Note that the definition of the quasiparticle energy
$\omega_{\mathbf{k}} \equiv |\phi^\prime_k|/|\phi_k|$,
oscillates even in the adiabatic limit, 
with the minimum and maximum values given by 
$\omega_{\mathbf{k},\rm min} 
   = \epsilon_{\mathbf{k}} (|\alpha_0|-|\beta_0|)/(|\alpha_0|+|\beta_0|)$
and $\omega_{\mathbf{k},\rm max} 
   = \epsilon_{\mathbf{k}}^2/\omega_{\mathbf{k},\rm min}$, respectively, 
such that $\omega_{\mathbf{k}}\neq \epsilon_{\mathbf{k}}$ in general.
This indicates that imposing instantaneous 
equality of the potential and kinetic energies
may not be appropriate in general situations. When particle number is understood
as  an average over the characteristic oscillation period however, 
imposing equality of the potential and kinetic energy may lead to 
a reasonable definition for the particle number density.
}. 
Hence, for the state~(\ref{WKB state})
our particle number definition~(\ref{particle-number:scalars:kin})
provides an upper limit for~(\ref{n:AartsBerges}).
This was to be expected, considering that 
Eq.~(\ref{particle-number:scalars:kin}) was derived 
in section~\ref{sec:Bogolyubov transformation} 
by an extremization procedure over the Bogolyubov
transformations~(\ref{Bogolyubov-transformation}).
We expect that a similar behaviour pertains in other situations.

\section{Fermions}
\label{Fermions}

Provided the fields are rescaled
as $a^{3/2}\psi\rightarrow \psi$ and the mass as $am\rightarrow m$, 
the fermionic Lagrangean reduces to the standard Minkowski form,
\begin{eqnarray}
 \sqrt{-g} {\cal L}_\psi \rightarrow 
 \bar{\psi}\, i\partial\!\!\!/\, \psi - \bar{\psi} (m_R + i \gamma^5 m_I) \psi
,
\label{Lagrangean:fermions}
\nonumber
\end{eqnarray}
where, for notational simplicity, we omitted the rescaling of the fields
and absorbed the scale factor in the mass term.
Note that the complex mass term
$  
m = m_{R}(\eta)+i m_{I}(\eta)
$
may induce CP-violation ({\it cf.} 
Ref.~\cite{KainulainenProkopecSchmidtWeinstock:2001+2002}).

The fermionic Wigner function, 
\begin{eqnarray}
   iS^<(k,x)
 = -\int d^4r {\rm e}^{ik\cdot r}
   \langle 0|\bar{\psi}(x-r/2)\psi(x+r/2)|0\rangle
\nonumber
\end{eqnarray}
satisfies the corresponding Dirac equation which, 
in the Wigner representation,
reads
\begin{equation}
\Bigl( {k}\!\!\!/\; + \frac{i}{2}\gamma^0 \partial_{t}
    - (m_{R}+i \gamma^5 m_{I})
       e^{-\frac{i}{2}\stackrel{\!\!\leftarrow}{\partial_{t}}\partial_{k_0}}
\Bigr) iS^{<} = 0
,
\label{S<}
\end{equation}
where $(i\gamma^0S^{<})^\dagger = i\gamma^0S^{<}$ is hermitean.
The helicity operator in the Weyl representation
%
$
\hat{h} = \hat{\mathbf{k}}\cdot\gamma^{0}
          \mbox{\boldmath{{$\gamma$}}}\gamma^{5}
$
%
commutes with the Dirac operator in (\ref{S<}), such that 
we can make the helicity block-diagonal ansatz for the Wigner function
({\it cf.} Ref.~\cite{KainulainenProkopecSchmidtWeinstock:2001+2002}) 
\begin{equation}
iS^{<} = \sum_{h=\pm}iS^{<}_{h}
\,,\qquad
-i\gamma_{0}S^{<}_{h}
  = \frac{1}{4}\bigl(\mathbbm{1}+h\hat{\mathbf{k}}\cdot
                     \mbox{\boldmath{{$\sigma$}}}
               \bigr)\otimes\rho^{a}g_{ah}
,
\label{S<:helicity-diagonal}
\end{equation}
where  $\hat{\mathbf{k}} = \mathbf{k}/|\mathbf{k}|$
and $\sigma^{a}$, $\rho^{a}$ ($a=0,1,2,3$) are the Pauli matrices.
Taking the traces of $\{\mathbbm{1},-h\gamma^i\gamma^5,-ih\gamma^i,-\gamma^5\}$ 
times the real part of~(\ref{S<}), and integrating
over $k_0$, yields the kinetic equations for the 0th momenta of $g_{ah}$,
\begin{eqnarray}
                                      \dot{f}_{0h} &=& 0
\label{f0eq}
\\
\dot{f}_{1h} + 2h|\mathbf{k}|f_{2h} - 2m_{I}f_{3h} &=& 0
\label{f1eq}
\nonumber\\
\dot{f}_{2h} - 2h|\mathbf{k}|f_{1h} + 2m_{R}f_{3h} &=& 0
\label{f2eq}
\nonumber\\
        \dot{f}_{3h} - 2m_{R}f_{2h} + 2m_{I}f_{1h} &=& 0
\,,
\label{f3eq}
\end{eqnarray}
where 
\begin{eqnarray}
 f_{0h} &\equiv& Tr\Big[(\mathbbm{1}P_h)\int\frac{dk_0}{2\pi}(-i\gamma^0S^<)\Big]
\nonumber\\ 
f_{1h}&\equiv& Tr\Big[
                     (-h\hat \mathbf{k}\cdot\mbox{\boldmath{{$\gamma$}}}\gamma^5
                     P_h)\int\frac{dk_0}{2\pi}(-i\gamma^0S^<)
                  \Big]
\nonumber\\ 
f_{2h} &\equiv& Tr\Big[
                       (-ih\hat\mathbf{k}\cdot\mbox{\boldmath{{$\gamma$}}}P_h)
                         \int\frac{dk_0}{2\pi}(-i\gamma^0S^<)
                  \Big]
\nonumber\\ 
f_{3h} &\equiv& Tr\Big[(-\gamma^5P_h)\int\frac{dk_0}{2\pi}(-i\gamma^0S^<)\Big]
\,,
\label{f_ha:definitions}
\end{eqnarray}
and $P_h = (1/2)[1+ h \hat\mathbf{k}\cdot\mbox{\boldmath{{$\gamma$}}}\gamma^5]$
denotes the helicity projector.
Eq.~(\ref{f0eq}) expresses the conservation of the Noether vector current.
The traces of the imaginary parts of~(\ref{S<})
decouple from~(\ref{f3eq}) at tree level,
and hence are of no importance for the analysis presented here. 
The moments $f_{ah}$ can be related to the positive and negative frequency
mode functions, $u_{h}(\mathbf{k},t)$ and 
$v_{h}(\mathbf{k},t) = - i\gamma^2(u_{h}(\mathbf{k},t))^{*}$,
respectively.
They form a basis for the Dirac field,
\begin{eqnarray}
\psi(x) \!=\! \frac 1V\sum\limits_{\mathbf{k}h}
          {\rm e}^{\!-i\mathbf{k}\cdot\mathbf{x}}
    \big(u_{h}a_{\mathbf{k}h}
       + v_{h}b_{-\mathbf{k}h}^{\dagger}
    \big)
,\quad\!\!
u_{h} \!=\! \biggl(\begin{array}{c}\!L_{h}\! \\ 
                              \!R_{h}\!
              \end{array}
        \biggr)\otimes \xi_{h}
,
\nonumber
\end{eqnarray}
where $\xi_{h}$ is the helicity two-eigenspinor,
$\hat h \xi_{h} = h \xi_{h}$. The Dirac equation then decomposes into
\begin{eqnarray}
i\partial_{0}L_{h}-h|\mathbf{k}|L_{h} &=& m_{R}R_{h}+im_{I}R_{h}
\nonumber\\
i\partial_{0}R_{h}+h|\mathbf{k}|R_{h} &=& m_{R}L_{h}-im_{I}L_{h}
.
\label{LhRh}
\end{eqnarray}
Note that these equations incorporate CP-violation and thus 
generalize the analysis of
Refs.~\cite{MamaevMostepanenkoFrolov:1976,ChungKolbRiottoTkachev:1999,PelosoSorbo:2000}.
Now, from~(\ref{LhRh}) one can derive~(\ref{f0eq})-(\ref{f3eq}) 
by multiplying with $L_{h}$ and $R_{h}$ and employing
\begin{eqnarray}
 f_{0h} &=& |L_{h}|^2+|R_{h}|^{2},
\quad\;
 f_{3h} = |R_{h}|^2-|L_{h}|^2
\nonumber\\
 f_{1h} &=& -2\Re(L_{h}R_{h}^{*}),
\qquad
 f_{2h} = 2\Im(L_{h}^{*}R_{h})
.
\label{f3asLR}
\end{eqnarray}
%
The Hamiltonian density reads
\begin{eqnarray}
H \!=\! \frac 1V\!\sum\limits_{\mathbf{k}h}\!
     \Bigl\{\Omega_{\mathbf{k}h}\bigl(a^{\dagger}_{\mathbf{k}h}a_{\mathbf{k}h}
  \!+\!  b^{\dagger}_{\!-\mathbf{k}h}b_{\!-\mathbf{k}h}\bigr)
  \!+\!(\Lambda_{\mathbf{k}h}b_{\!-\mathbf{k}h}a_{\mathbf{k}h}
  \!+\! {\rm h.c.})\!
     \Bigr\}
\nonumber
\end{eqnarray}
where 
\begin{eqnarray}
\Omega_{\mathbf{k}h} 
  &=& h{k}
      \left(|L_{h}|^{2}-|R_{h}|^{2}
      \right)
   + mL_{h}^{*}R_{h}
   + m^{*}L_{h}R_{h}^{*}
\nonumber\\
\Lambda_{\mathbf{k}h}
  &=& 2{k}L_{h}R_{h}
        - hm^{*}L_{h}^{2}+hmR_{h}^{2}
,\quad
\label{hamiltonian:fermions}
\end{eqnarray}
with $\{\hat{a}_{\mathbf{k}h},\hat{a}_{\mathbf{k'}h'}^{\dagger}\}
       = \delta_{h,h'}\delta_{\mathbf{k},\mathbf{k'}}$,
 $\{\hat{b}_{\mathbf{k}h},\hat{b}_{\mathbf{k'}h'}^{\dagger}\}
       = \delta_{h,h'}\delta_{\mathbf{k},\mathbf{k'}}$.
We now use the Bogolyubov transformation
\begin{eqnarray}
\left(\begin{array}{c}
            \hat{a}_{\mathbf{k}h}\\
            \hat{b}^{\dagger}_{-\mathbf{k}h}
      \end{array}\right)
 \!=\!
\left(\!\begin{array}{cc}
            \alpha_{{k}h}
          & \beta_{{k}h}\\
            -\beta_{{k}h}^{*}
          & \alpha_{{k}h}^{*}
      \end{array}\!\right)
\left(\!\begin{array}{c}
            a_{\mathbf{k}h}\\
            b_{-\mathbf{k}h}^{\dagger}
      \end{array}\!\right)
,
\nonumber
\end{eqnarray}
to diagonalize the Hamiltonian, 
where $\alpha_{{k}h}$ and $\beta_{{k}h}$ are
\begin{equation}
 \frac{1}{2}\bigg(
         \bigg|\frac{\alpha_{{k}h}}{\beta_{{k}h}}\bigg|
       - \bigg|\frac{\beta_{{k}h}}{\alpha_{{k}h}}\bigg|
     \bigg)
 = \frac{\Omega_{\mathbf{k}h}}{|\Lambda_{\mathbf{k}h}|}
,\quad
      |\alpha_{{k}h}|^{2}+|\beta_{{k}h}|^{2}  = 1
\label{Bogolyubov-transformation:fermions}
,
\end{equation}
leading to the particle number density on phase space,
\begin{eqnarray}
 n_{\mathbf{k}h} &=& 
                    |\beta_{{k}h}|^{2}
               = \frac{1}{2} - \frac{\Omega_{\mathbf{k}h}}{2\omega_{\mathbf{k}}}
\,,
\label{nk-fermionsBog}
\end{eqnarray}
where now $\omega_{\mathbf{k}} =\sqrt{\mathbf{k}^2+|m|^2}$.

To construct the initial mode functions in the adiabatic domain,
$\eta\rightarrow -\infty$, we use the positive frequency solution
and its charge conjugate,
\begin{eqnarray}
\psi_{\mathbf{k}}\rightarrow 
  \left(\begin{array}{c}\alpha_0L_{h}^{+} + \beta_0L_{h}^{-}\\
                        \alpha_0R_{h}^{+} + \beta_0R_{h}^{-}
        \end{array}
  \right)
,\qquad
|\alpha_0|^{2} + |\beta_0|^{2}=1
.
\nonumber
\end{eqnarray}
From the Dirac equation under adiabatic conditions it follows
\begin{eqnarray}
L_{h}^{+} &=& \sqrt{\frac{\omega_{\mathbf{k}}+h{k}}{2\omega_{\mathbf{k}}}}
,
\qquad\qquad
L_{h}^{-} = -i\frac{m}{|m|}
              \sqrt{\frac{\omega_{\mathbf{k}}-h{k}}{2\omega_{\mathbf{k}}}}
\nonumber\\
R_{h}^{+} &=& \frac{m^{*}}
                   {\sqrt{2\omega_{\mathbf{k}}(\omega_{\mathbf{k}}+h{k})}}
,
\qquad
R_{h}^{-} = i\frac{|m|}{\sqrt{2\omega_{\mathbf{k}}(\omega_{\mathbf{k}}-h{k})}}
\,.
\nonumber
\end{eqnarray}
These mode functions correspond to an initial
particle number $n_{\mathbf{k}}^{(0)} = |\beta_0|^2$. 
We now make use of~(\ref{f3asLR})
to express $\Omega_{\mathbf{k}h}$ in terms of the Wigner functions,
\begin{equation}
\Omega_{\mathbf{k}h} = - (h{k}f_{3h} + m_{R}f_{1h} + m_{I}f_{2h})
,
\end{equation}
which implies our main result for fermions,
\begin{equation}
 n_{\mathbf{k}h} = \frac{1}{2\omega_{\mathbf{k}}}\left( h{k}f_{3h}
                                           + m_{R}f_{1h}
                                           + m_{I}f_{2h}
                                      \right)
                + \frac{1}{2}
\,.
\label{particle-number:fermions:kin}
\end{equation}
Note that in the limit $m\rightarrow 0$, this expression 
reduces to the phase space density of axial particles. Moreover,
$0\leq n_{\mathbf{k}h} \equiv |\beta_{kh}|^2 = 1- |\alpha_{kh}|^2 \leq 1$
(see Eqs.~(\ref{Bogolyubov-transformation:fermions}-\ref{nk-fermionsBog})).

As an application of Eq.~(\ref{particle-number:fermions:kin}) 
we consider particle production at 
preheating~\cite{PelosoSorbo:2000,KofmanLindeStarobinsky:1997}, 
in which the fermionic mass is generated by an oscillating inflaton
condensate. Assuming that the inflaton oscillates as a cosine function
results in a fermion production shown in 
figure~\ref{figure 1}. Observe that, even for a relatively small imaginary
(pseudoscalar) mass term, particle production of the opposite helicity 
states is completely different, implying a nonperturbative 
enhancement of a CP-violating particle density,
$n_{\mathbf{k}+}-n_{\mathbf{k}-}$, which may be of
relevance for baryogenesis.
\begin{figure}[htbp]
\begin{center}
\epsfig{file=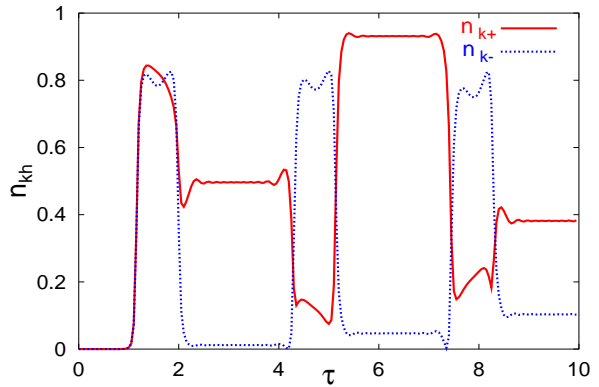, height=2.2in,width=3.3in}
\end{center}
\vskip -0.35in
\lbfig{figure 1}
\caption[fig1]{%
\small
The number of produced fermions as a function of time with
helicity $h=+$ ({\it solid}\,) and $h=-$ ({\it dotted}\,), 
mass $m/\omega_I=10+15\cos(2\tau)-i\sin(2\tau)$, $|\mathbf{k}|=\omega_I$,
$\tau = \omega_I t$, where $\omega_I$ denotes the frequency of the inflaton 
oscillations.
}
\vskip -0.15in
\end{figure}

When applied to thermal equilibrium, where 
({\it cf.} Ref.~\cite{LeBellac:1996})
\begin{eqnarray}
iS^{<} = - ({k}\!\!\!/\, + m_R - i\gamma_5m_I)\delta(k^{2}\!-\!|m|^{2})
           \frac{2\pi\textnormal{sign}(k_{0})}{{\rm e}^{\beta k_{0}}+1}
,
\label{equilibrium:fermions}
\end{eqnarray}
we find
\begin{eqnarray} 
f_{0h} &=& 1,\nonumber\\
f_{1h} &=& (2m_R/\omega_{\mathbf{k}})
          [\{\exp(\beta\omega_{\mathbf{k}})+1\}^{-1} - 1/2],\nonumber\\
f_{2h} &=& (2m_I/\omega_{\mathbf{k}})
          [\{\exp(\beta\omega_{\mathbf{k}})+1\}^{-1} - 1/2],\nonumber\\
f_{3h} &=& (2hk/\omega_{\mathbf{k}})
          [\{\exp(\beta\omega_{\mathbf{k}})+1\}^{-1} - 1/2],\nonumber
\end{eqnarray}
such that Eq.~(\ref{particle-number:fermions:kin})
yields the Fermi-Dirac distribution,
\begin{equation}
 n_{\mathbf{k}h} = {1}/({{\rm e}^{\beta\omega_{\mathbf{k}}}+1}).\nonumber
\end{equation}

\section{Multiflavour Case}

We now generalize the definition of particle number in terms of
two-point functions to the case of several species, mixing through a mass 
matrix. While in the single flavour case always an equal number of particles
and antiparticles is produced, we will here encounter the creation of
a charge asymmetry when the mass matrix is nonsymmetric. Because of this 
charge violation,
the orthogonality of particle modes with respect to antiparticle modes is not
preserved under time evolution, and it is thus impossible to expand the
field operators in terms of an orthogonal basis.

Hence, the use of the basis-independent two-point functions is advantageous.
We can either calculate the time evolution of the system in terms of these
quantities or measure them, since they correspond to physical charge- and
current densities. When finally the mass matrix is diagonal and only
adiabatically slowly evolving, there exists a well-defined basis, in terms of 
which the Hamiltonian is diagonal. We use this basis to define the particle
number operators and construct their expectation values out of the two-point
functions. 

\subsection{Fermions}
Since Dirac spinors naturally include particle and antiparticle modes,
we first discuss here the fermionic case.
We decompose the mass matrix $M$ into a hermitean and an antihermitean part,
\begin{equation}
M_H=\frac{1}{2}(M+M^\dagger),\quad M_A=\frac{1}{2i}(M-M^\dagger),
\end{equation}
such that the Dirac equation reads
\begin{equation}
\left[i{\partial}\!\!\!/\ -M_H-i\gamma^5 M^A\right]_{ij}\psi_j=0\label{muflaDirac}.
\end{equation}
One can then attempt to proceed as in the single flavour case and
to construct the field operators as
\begin{eqnarray}
\!\!\!\!\psi_i\!(x)\!\!\!&=&\!\!\!\sum_{\mathbf{k}\tilde h j}\!
          \frac{ {\rm e}^{\textnormal - i\mathbf{k}\cdot\mathbf{x}}}{V}\!\!
\left[
U_{\tilde h\, i\!j}(\mathbf{k},t)a_{\tilde h\, j}(\mathbf{k})
\!+\!V_{\tilde h\, i\!j}(\mathbf{k},t)b^{\dagger}_{\tilde h\, j}(\textnormal{-}\mathbf{k})\!
\right]\label{psiexp}\\
\!\!\!\!\psi_i^{\dagger}\!(x)\!\!\!&=&\!\!\!\sum_{\mathbf{k}\tilde h j}\!
                 \frac{{\rm e}^{\textnormal -i\mathbf{k}\cdot\mathbf{x}}}{V}\!\!
\left[
a_{\tilde h\, j}^{\dagger}(\mathbf{k})U_{\tilde h\, ji}^{\dagger}(\mathbf{k},t)
\!+\!b_{\tilde h\, j}(\textnormal{-}\mathbf{k})V_{\tilde h\, ji}^{\dagger}(\mathbf{k},t)\!
\right],\nonumber
\end{eqnarray}
with the mode function
\begin{equation}
U_{h\, ij}=\biggl(\begin{array}{c}\!L_{h\,ij}\! \\ 
                           \!R_{h\,ij}\!
              \end{array}
        \biggr)\otimes \xi_{h}
\end{equation}
and its charge conjugate
\begin{equation}
V_{h\, ij}=-i\gamma^2(U_{h\, ij})^*=
CU_{h\, ij}C^{-1}=\biggl(\begin{array}{c}-h\!R_{h\,ij}^{*}\! \\ 
                           h\!L_{h\,ij}^{*}\!
              \end{array}
        \biggr)\otimes \xi_{-h}.
\end{equation}

This procedure however fails, when $M$ is not a symmetric matrix, which
can easily be seen by plugging $U_{h\, ij}$ into Eq.~(\ref{muflaDirac})
\begin{eqnarray}
\left\{i\partial_0-h|\mathbf{k}|\right\}\mathbbm{1}_{il}L_{h\, ij}&=&
M^H_{il}R_{h\, lj}+iM^A_{il}R_{h\, lj}\label{Ueq}\nonumber\\
\left\{i\partial_0+h|\mathbf{k}|\right\}\mathbbm{1}_{il}R_{h\, ij}&=&
M^H_{il}L_{h\, lj}-iM^A_{il}L_{h\, lj}
\end{eqnarray}
and $V_{h\, ij}$, respectively,
\begin{eqnarray}
\left\{i\partial_0-h|\mathbf{k}|\right\}\mathbbm{1}_{il}L_{h\, ij}&=&
M^{H*}_{il}R_{h\, lj}+iM^{A*}_{il}R_{h\, lj}\label{Veq}\nonumber\\
\left\{i\partial_0+h|\mathbf{k}|\right\}\mathbbm{1}_{il}R_{h\, ij}&=&
M^{H*}_{il}L_{h\, lj}-iM^{A*}_{il}L_{h\, lj}
\,,
\end{eqnarray}
where summation over the repeated index $l$ is implied.

Obviously, when $M$ is not symmetric, Eqs.~(\ref{Ueq}) and~(\ref{Veq}) are
inconsistent.
In particular, for nonsymmetric $M$, the orthogonality condition
\begin{equation}
U_{r\,il}^\dagger V_{s\,lj}=0,\label{ortho}
\end{equation}
is not preserved at all times, and hence, the expansion of the
field operators~(\ref{psiexp}) is not suitable. 
This complication can however lead to the generation of a net charge
stored in the produced particles, because the operation of charge
conjugation becomes time dependent, an effect which may be of
relevance for baryogenesis~\cite{Garbrecht:2003mn}.

The construction of an appropriate Bogolyubov transformation for the
case of a symmetric mass matrix is
discussed in Ref.~\cite{Nilles:2001fg}. In comparison with the
single flavour case this procedure is fairly complicated. 
For the general case, we therefore refrain from a computation of a Bogolyubov
transformation and the time evolution of Heisenberg creation and
annihilation operators.

It is more convenient to calculate the time evolution of the initial state
in terms of two point functions.
We straightforwardly generalize the formalism for the
single-flavour Wigner funtions to the multiflavour case by defining
\begin{eqnarray}
   iS^<_{ij}(k,x)
 = -\int d^4r {\rm e}^{ik\cdot r}
   \langle \bar{\psi}_{j}(x-r/2)\psi_{i}(x+r/2) \rangle
,
\end{eqnarray}
wher $a$, $b$ are flavour indices.
These obey the equation of motion
\begin{equation}
\Bigl( {k}\!\!\!/\; + \frac{i}{2}\gamma^0 \partial_{t}
    - (M_{H}+i \gamma^5 M_{A})
      {\rm e}^{-\frac{i}{2}\stackrel{\!\!\leftarrow}{\partial_{t}}\partial_{k_0}}
\Bigr)_{\!\!il} iS_{lj}^{<} = 0
\label{DiracEq}
\,.
\end{equation}
As described for the single flavour case in section~\ref{Fermions},
this can be simplified and yields
\begin{eqnarray}
\dot{f}_{0h} + i\left[M_H,f_{1h}\right] +i\left[M_A,f_{2h}\right] &=& 0
\label{f0eqN}\nonumber
\\
\dot{f}_{1h} + 2h|\mathbf{k}|f_{2h} + i\left[M_H,f_{0h}\right] -\left\{M_A,f_{3h}\right\}  &=& 0
\label{f1eqN}
\nonumber\\
\dot{f}_{2h} - 2h|\mathbf{k}|f_{1h}  + \left\{M_H,f_{3h}\right\} +i\left[M_A,f_{0h}\right] &=& 0
\label{f2eqN}
\nonumber\\
\dot{f}_{3h} - \left\{M_H,f_{2h}\right\} +\left\{M_A,f_{1h}\right\}  &=& 0
.
\end{eqnarray}
As already noted in~\cite{Garbrecht:2003mn}, we can infer from these equations 
as a necessary condition for the nonconservation of the charge density
$f_{0h}$,
that $M$ must not be symmetric, in accordance with our discussion above.

Now assume, that after some time evolution, $M$ has become symmetric
and slowly varying. Then, it is possible to expand the field operators as
in Eq.~(\ref{psiexp}) and to define the expectation values of the
number of particles
\begin{equation}
n^+_{\mathbf{k}h\,i}=\langle a^{\dagger}_{h\,i}(\mathbf{k})a_{h\,i}(\mathbf{k})
\rangle\nonumber
\end{equation}
and antiparticles
\begin{equation}
n^-_{\mathbf{k}h\,i}=\langle b^{\dagger}_{h\,i}(\mathbf{k})b_{h\,i}(\mathbf{k})
\rangle.\nonumber
\end{equation}

Moreover, we choose this basis such that the
Hamilton operator is diagonal and reads
\begin{eqnarray}
H=\frac{1}{V}\sum\limits_{\mathbf{k}hij}&\Big(&
h|\mathbf{k}|L_h^\dagger L_h+L_h^\dagger \left[M^H+iM^A\right]R_h
\label{hdiag}\nonumber\\
&-&h|\mathbf{k}|R_h^\dagger R_h+R_h^\dagger \left[M^H-iM^A\right]L_h
\Big)_{ij}\nonumber\\
&\times&\Big(
a_{hi}^\dagger (\mathbf{k})a_{hj}(\mathbf{k})-b_{hi}(\mathbf{k})b_{hj}^\dagger (\mathbf{k})
\Big).
\end{eqnarray}


We can now also express the functions $f^{ij}_{\mu h}$ employing this
basis. Explicitly, they read
\begin{eqnarray}
f^{ij}_{0h}(x,\mathbf{k}) \!\!= \!\! - \!\!\int d^4r\, {\rm e}^{ik\cdot r}
   \langle \bar{\psi}_{hj}(x-r/2) \gamma^0 \psi_{hi}(x+r/2)\rangle
\nonumber\\
 \!\!= \!\!\left({L^{il}_{h}}^* L^{jl'}_{h}+{R^{il}_{h}}^{*}R^{jl'}_h\right)
\times\left\langle
a_{hl'}^\dagger (\mathbf{k})a_{hl}(\mathbf{k})
     + b_{hl'}(\mathbf{k})b_{hl}^\dagger (\mathbf{k})
\right\rangle\nonumber
\\
f^{ij}_{1h}(x,\mathbf{k})  \!\!= \!\! - \!\!\int d^4r\, {\rm e}^{ik\cdot r}
   \langle \bar{\psi}_{hj}(x-r/2)  \psi_{hi}(x+r/2)\rangle
\nonumber\\
 \!\!= \!\!-2\Re\left(L^{il}_{h}{R^{jl'}_{h}}^{*}\right)
\times\left\langle
a_{hl'}^\dagger (\mathbf{k})a_{hl}(\mathbf{k})
      +b_{hl'}(\mathbf{k})b_{hl}^\dagger (\mathbf{k})
\right\rangle,\nonumber
\\
f^{ij}_{2h}(x,\mathbf{k}) \!\! = \!\! - \!\!\int d^4r\, {\rm e}^{ik\cdot r}
   \langle \bar{\psi}_{hj}(x-r/2) (-i\gamma^5)\psi_{hi}(x+r/2)\rangle
\nonumber\\
 \!\!= \!\! 2\Im\left({L^{il}_{h}}^{*}R^{jl'}_{h}\right)
\times\left\langle
a_{hl'}^\dagger (\mathbf{k})a_{hl}(\mathbf{k})
  +b_{hl'}(\mathbf{k})b_{hl}^\dagger (\mathbf{k})
\right\rangle\nonumber
,
\\
f^{ij}_{3h}(x,\mathbf{k}) \!\! = \!\! - \!\!\int d^4r\, {\rm e}^{ik\cdot r}
   \langle 
           \bar{\psi}_{hj}(x-r/2) \gamma^0\gamma^5 \psi_{hi}(x+r/2)
   \rangle
\nonumber\\
 \!\!= \!\! \left({L^{il}_{h}}^{*}L^{jl'}_{h}-{R^{il}_{h}}^{*}R^{jl'}_h\right)
\times\left\langle
a_{hl'}^\dagger (\mathbf{k})a_{hl}(\mathbf{k})
      +b_{hl'}(\mathbf{k})b_{hl}^\dagger (\mathbf{k})
\right\rangle\nonumber
.
\end{eqnarray}

By comparing with the expression~(\ref{hdiag}), we obtain
\begin{equation}
\langle H\rangle=-\frac{1}{V}\sum\limits_{\mathbf{k}hi}
h|\mathbf{k}|f_{3h\,ii}+M^H_{ii}f_{1h,ii}+M^A_{ii}f_{2h\,ii}.
\end{equation}

We define $\omega_{i}(\mathbf{k})=(\mathbf{k}^2+|M_{ii}|^2)^{1/2}$, and
since we assumed diagonality of the Hamiltonian, this has to equal
\begin{eqnarray}
\langle H\rangle&=&\frac{1}{V}\sum\limits_{\mathbf{k}hi}
\omega_i(\mathbf{k})\langle a^{\dagger}_{h\,i}(\mathbf{k})a_{h\,i}(\mathbf{k})
- b_{h\,i}(\mathbf{k})b^{\dagger}_{h\,i}\rangle\nonumber\\
&=&\frac{1}{V}\sum\limits_{\mathbf{k}hi}
\omega_i(\mathbf{k}) \left(n^+_{\mathbf{k}hi} + n^-_{\mathbf{k}h\,i} -1\right),
\end{eqnarray}
while the charge is
\begin{equation}
f_{0h\,ii}=\langle a^{\dagger}_{h\,i}(\mathbf{k})a_{h\,i}(\mathbf{k})
+ b_{h\,i}(\mathbf{k})b^{\dagger}_{h\,i}\rangle
=n^+_{\mathbf{k}h\,i} - n^-_{\mathbf{k}h\,i} + 1.
\end{equation}
We thus find the following generalization of~(\ref{particle-number:fermions:kin})
\begin{eqnarray}
 \!\!n^+_{\mathbf{k}hi} \!\!\!&=&\! \!\!\frac{
                              h|\mathbf{k}|f_{3hii}
                                    \!+\!   M^H_{ii}f_{1hii}
                                   \! +\!   M^A_{ii}f_{2hii}
                                  }{2\omega_i(\mathbf{k})}
                   \!+\!   \frac{1}{2}f_{0hii}
\label{nk:mixing fermions+}
\\
 \!\!n^-_{\mathbf{k}hi} \!\!\!&=&\! \!\! \frac{
                                 h|\mathbf{k}|f_{3hii}
                                   \! +\!   M^H_{ii}f_{1hii}
                                    \!+\!   M^A_{ii}f_{2hii}
                                  }{2\omega_i(\mathbf{k})}
                     \!-\!   \frac{1}{2}f_{0hii}\!
                     +\!  1\!
\,,
\label{nk:mixing fermions-}
\end{eqnarray}
which is of course the anticipated result, since the number of particles
is just the half of the total particle number (particles plus antiparticles)
plus half of the total charge (particles minus antiparticles). 

\subsection{Scalars}

Consider now a complex scalar field $\Phi_i$ describing $N$ flavours,
which we expand into its hermitean and antihermitean parts as follows, 
\begin{equation}
\Phi_i=\frac{1}{\sqrt{2}}(\Phi_i^1+i\Phi_i^2),
\end{equation}
such that the multiflavour field operator is
\begin{eqnarray}
\Phi_i&=&\frac{\varphi_i}{a}
=\frac{1}{aV}\sum\limits_\mathbf{k}
e^{-i\mathbf{k}\cdot\mathbf{x}}\nonumber\\
&\times&\Big(\!
\varphi^1_{ij}(\mathbf{k},\eta)a^1_j(\mathbf{k})
+i\varphi^2_{ij}(\mathbf{k},\eta)a^2_j(\mathbf{k})\nonumber\\
&+&\!\!\varphi^{1\dagger}_{ij}(\textnormal-\mathbf{k},\eta)a^{1\dagger}_j(\textnormal-\mathbf{k})
+i\varphi^{2\dagger}_{ij}(\textnormal-\mathbf{k},\eta)a^{2\dagger}_j(\textnormal-\mathbf{k})
\Big),
\end{eqnarray}
where the rescaled fields obey the generalized Klein-Gordon equation
\begin{equation}
\left\{\partial_\eta^2+\mathbf{k}^2+M^2-
\frac{a''}{a}\right\}_{il}\varphi^\alpha_{lj}=0.
\end{equation}
Note that this is independent of whether $\alpha=1$ or $\alpha=2$, 
which is just as
in the fermionic case, where the functions $U$ and $V$ both satisfy the
Dirac equation.
The individual components $\Phi_i^1$ and $\Phi_i^2$ are imposed to be hermitean.
Therefore,
$\sum_j  (\varphi_{ij}(\mathbf{k})+\varphi_{ij}(\mathbf{-k}))$
has to be real, which can in general be satisfied only if $M^2$ is
real or, more precisely, real symmetric.

Let us therefore assume again, that we are in a final state with
diagonal and only nonadiabatically varying $M$. We define
\begin{eqnarray}
a(\mathbf{k})&=&\frac{1}{\sqrt{2}}\left[a^1(\mathbf{k})+ia^2(\mathbf{k})\right]
\quad\textnormal{and}\nonumber\\
b(\mathbf{k})&=&\frac{1}{\sqrt{2}}\left[a^1(\mathbf{k})-ia^2(\mathbf{k})\right].
\end{eqnarray}
Then, we find the charge operator to be
\begin{equation}
Q_i(\mathbf{k})=\langle a_i^\dagger(\mathbf{k})a_i(\mathbf{k})
-b_i^\dagger(\mathbf{k})b_i(\mathbf{k})\rangle\label{QScalar}
\end{equation}
and the Hamiltonian
\begin{equation}
H=\frac{1}{V}\sum\limits_\mathbf{k}\Omega_i(\mathbf{k})\left(
a_i^\dagger(\mathbf{k})a_i(\mathbf{k})
+b_i^\dagger(\mathbf{k})b_i(\mathbf{k})+1
\right),
\end{equation}
where
$\Omega_i(\mathbf{k})
    =\left|\varphi_i'(\mathbf{k})-(a'/a)\varphi_{i}(\mathbf{k})
        \right|^2
     + \omega_i(\mathbf{k})^2 \left|\varphi_i(\mathbf{k})\right|^2$,
$\omega_i^2(\mathbf{k})=\mathbf{k}^2+a^2M^2_{ii}$.

We define the multiflavour Wightman function as
\begin{equation}
i\bar{G}^<_{ij}=\langle\varphi^\dagger_j(u)\varphi_i(v)\rangle
\end{equation}
and adapt straightforwardly the definition of the momenta from the
single flavour case. These then satisfy the system of equations
\begin{eqnarray}
\frac{1}{4}f_{0}'\!'-\!f_{2}\!+\!\frac{1}{2}\left\{M^2,f_{0}\right\}
\!+\!\left(\mathbf{k}^2-\frac{a''}{a}\right)f_{0\!\!\!}&=&\!\!\!0\nonumber\\
f_{1}\!'-\!\frac{i}{2}\left[M^2,f_{0}\right] \!\!\!&=\!\!\!&0\nonumber\\
f_{2}'\!-\!\frac{i}{2}\left[M^2,f_{1}\right]\!-\!\frac{1}{4}
\left\{(\mathbf{k}^2+M^2-a''/a)',f_{0}\right\}\!\!\!&=\!\!\!&0.
\end{eqnarray}
We find $Q_i(\mathbf{k})=f_{1\,ii}(x,\mathbf{k})+1$, 
which is also in accordance with 
the $U(1)$-Noether charge. Together with the
identities~(\ref{dotg}), this leads us to
\begin{eqnarray}
n^+_{\mathbf{k}\,i}\!&=&\!\omega_if_{0\,ii}
                    \!+\! \frac{f_{0\,ii}''}{4\omega_i}
\!-\!\frac{1}{2\omega_i}
                      \frac{d}{d\eta}\left(\frac{a'}{a}f_{0\,ii}\right)
+\frac{1}{2}f_{1\,ii}
\label{nk:mixing scalars+}
\\
n^-_{\mathbf{k}\,i}\!&=&\!\omega_if_{0\,ii}
                    \!+\! \frac{f_{0\,ii}''}{4\omega_i}
\!-\!\frac{1}{2\omega_i}
                      \frac{d}{d\eta}\!\left(\frac{a'}{a}f_{0\,ii}\right)
\!-\!\frac{1}{2}f_{1\,ii}-\!1,
\label{nk:mixing scalars-}
\end{eqnarray}
where $n^+_{\mathbf{k}\,i}$ is the number of particles,
 $n^-_{\mathbf{k}\,i}$ the number of antiparticles,
and the same simple interpretion as in the fermionic case applies.

\section{Discussion}

 We have derived general expressions for the particle number densities 
on phase space for single scalars~(\ref{particle-number:scalars:kin}) 
and fermions~(\ref{particle-number:fermions:kin})
in terms of the appropriate Wigner functions. We have then generalized our
analysis to the case of mixing 
scalars~(\ref{nk:mixing scalars+}--\ref{nk:mixing scalars-}) and 
fermions~(\ref{nk:mixing fermions+}--\ref{nk:mixing fermions-}).
All of these expressions are positive, and moreover, 
the number of fermions is bounded from
above by {\it unity}, as required by the Pauli principle. 
In order to incorporate the effect of self-energy
into~(\ref{particle-number:scalars:kin}) 
and~(\ref{particle-number:fermions:kin}), one needs to include  
this correction into the dispersion relation,
$\omega=\omega(\mathbf{k},x)\rightarrow
   \omega + \Sigma_H(\mathbf{k},x)$, where 
$\Sigma_H(\mathbf{k},x) 
  \equiv \int[dk_0/(2\pi)](1/2)[\Sigma^r(k,x)+\Sigma^a(k,x)]$,
and $\Sigma^r$ and $\Sigma^a$ denote the retarded and advanced self-energies,
respectively~\cite{ProkopecSchmidtWeinstock:2003}. When the single particle
picture breaks down it is not clear whether a sensible definition of particle 
number can be constructed. Our analysis can be quite straightforwardly
extended to include (time-varying) gauge fields by coupling them canonically
to scalars and fermions.

The kinetic theory definition of the particle number is of course and
by construction identical
with the definition in terms of Bogolyubov transformations. The number
of individual particles is the total energy of the system divided by
the energy of an individual particle. Taking the point of view of
kinetic theory proves advantageous when considering the multiflavour case
or  statistical systems, such as the thermal equilibrium.

While the fermionic particle number 
definition~(\ref{particle-number:fermions:kin}) is 
generally applicable, the scalar 
one~(\ref{particle-number:scalars:kin}) fails however when 
$\omega_{\mathbf{k}}^2 = {\mathbf{k}}^2 + a^2 m_\phi^2 < 0$, 
which can happen at phase transitions.  
Then $\Omega_{\mathbf{k}} < |\Lambda_{\mathbf{k}}|$ 
in (\ref{Hamiltonian:bosons}), 
and the Bogolyubov transformation~(\ref{Bogolyubov-transformation}) 
does not have a solution. 
Nevertheless, even in this case, the energy density on phase 
space $\Omega_{\mathbf{k}}$ in
Eq.~(\ref{Omega:kinetic-theory}) is well defined,
and should be considered as a fundamental quantity of kinetic theory.
Another important quantity is 
$\Lambda^*_{\mathbf{k}} = \langle \mathbf{k},-\mathbf{k} | H | 0 \rangle $,
the transition amplitude for particle pair creation with the momenta
$\{\mathbf{k},-\mathbf{k}\}$; and likewise
$\Lambda_{\mathbf{k}}$ is the transition amplitude for pair annihilation. 
The appropriate description in this case is in terms of squeezed states.
For an account of the inverted harmonic oscillator in terms of squeezed states 
see {\it eg.} Ref.~\cite{AlbrechtFerreiraJoyceProkopec:1992}.

Our definition of particle number can be used for studies 
of quantum-to-classical transition, decoherence and entropy calculations
of {\it eg.} cosmological 
perturbations~\cite{BrandenbergerMukhanovProkopec:1992,AlbrechtFerreiraJoyceProkopec:1992,PolarskiStarobinsky:1995}.
Moreover, 
when suitably normalized, the particle density $n_{\mathbf{k}}$ can be
used to define a density matrix on phase space, 
$\varrho_{\mathbf{k}} = n_{\mathbf{k}}/\sum_{\mathbf{k'}}  n_{\mathbf{k'}}$.

In the derivation of our results, we considered pure quantum states, yet
showed explicitly their applicability to thermal states. More generally,
our definitions are valid if one
requires the density matrix $\varrho$ to satisfy
$
\langle a_{\mathbf{k}} a_{\mathbf{k}}\rangle_{\varrho}
 = \langle a^\dagger_{\mathbf{k}} a^\dagger_{\mathbf{k}}\rangle_{\varrho}
 = 0. 
$
These relations hold {\it e.g.} for eigen\-states of the particle number
operator $\hat N_{\mathbf{k}} \equiv a^\dagger_{\mathbf{k}} a_{\mathbf{k}}$,
and, as pointed out in Ref.~\cite{Kandrup:1988}, for random phase states,
a special case of which is the canonical ensemble.
States of this kind can be treated as a linear superposition of the particle
number eigenstates which we considered above.

Finally, we note that after the first version of this article appeared,
an out-of-equilibrium investigation of the dynamics of chiral fermions
coupled to scalars was studied in Ref.~\cite{Berges:2002wr}. 
In order to show that at late times the system thermalizes to the Fermi-Dirac 
equilibrium, the authors used a particle number definition, which can be
in our notation written as
\begin{equation}
 \tilde{n}_{\mathbf{k}}=\frac{1}{2}\sum_{h=\pm}\tilde{n}_{\mathbf{k}h}
\,,\qquad
 \tilde{n}_{\mathbf{k}h}=\frac{1}{2}\left(1+hf_{3h}\right)
\,.
\label{fermions:alternative definition}
\end{equation}
This definition corresponds to the massless fermion limit, 
$m\rightarrow 0$, of our definition~(\ref{particle-number:fermions:kin}).

\section*{Acknowledgements} 

We would like to acknowledge discussions with J\"urgen Berges, 
Dietrich B\"odeker, Kimmo Kainulainen and Julien Serreau.

%
%

\end{document}